# Quantification and Mapping of Elastic Strains in Ferroelectric [BaZrO$_3$]$_{x\Lambda}$/[BaTiO$_3$]$_{(1-x)\Lambda}$ Superlattices

J. Belhadi[1], F. Ravaux[2], H. Bouyanfif[1], M. Jouiad[1,a)] and M. El Marssi[1,a)]

[1]*Laboratory of Physics of Condensed Mater (LPMC), University of Picardie Jules Verne, Amiens, 80039, France.*

[2]*Materials Science department, Masdar Institute at Khalifa University, 54224, Abu Dhabi, United Arab Emirates.*

**Abstract**

We report on quantification and elastic strain mapping in two artificial [BaZrO$_3$]$_{x\Lambda}$/[BaTiO$_3$]$_{(1-x)\Lambda}$ (BZ$_{x\Lambda}$/BT$_{(1-x)\Lambda}$) superlattices having periods of $\Lambda$=6.6 nm and $\Lambda$ = 11 nm respectively, grown on (001) SrTiO$_3$ single crystal substrate by pulsed laser deposition technique. The methodology consists of a combination of high-resolution scanning transmission electron microscopy and nanobeam electron diffraction associated with dedicated algorithm for diffraction patterns processing originally developed for semiconductors to record the strains at atomic scale. Both in-plane and out-of-plane elastic strains were then determined at 2 nm spatial resolution and their average values were used to map the strains along and transverse to the epitaxial growth direction of both samples to determine its variation along several BZ/BT interfaces. In addition, the variation of the width of the inter-diffusion BT/BZ interfaces and intermixing between different layers are estimated. The obtained width average value measured in these inter-diffusion interfaces vary from 8 to 12% and from 9 to 11% for both superlattices having $\Lambda$ = 6.6 nm and $\Lambda$ = 11 nm respectively. These inter-diffusion interfaces and the inherent elastic strains due to the confined layers of the superlattices are known to be the most important parameters, responsible of the change in their functional properties.

[a]Authors to whom correspondence should be addressed. Electronic mails: mimoun.elmarssi@u-picardie.fr & Mustapha.jouiad@u-picardie.fr

# 1. Introduction

Superlattice (SL) structures composed of alternating complex oxide layers are fascinating materials that have attracted an increasing attention in the recent years. SLs are indeed ideal platforms to investigate dimensionality effects and interlayers coupling (elastic, electric and/or magnetic) in complex oxides with remarkable properties (ferromagnetism, superconductivity, ferroelectric). The different degrees of freedom (number of bilayers, constituents of the bilayer, relative proportion of each constituents in the bilayer) available in such platform allow a quasi-infinite possibility to investigate confinement, competing orders but also the design of new nanomaterials with emerging functional properties [1–4]. SLs hold different physical properties compared to those known in bulk materials or in conventional single films and new properties were recently discovered such as interfacial ferromagnetism, presence of vortex-array topology or induced ferroelectricity and/or magnetism in materials neither ferroelectric nor magnetic in their unstrained bulk state [4–7]. In addition, the functional properties of these modulated structures can be tuned and/or enhanced by engineering the elastic strains and the interfaces between different layers since many parameters can be varied in SLs, namely the nature of materials, modulation period and layer thickness of alternating compounds. Especially, the epitaxial strain is considered as one of the major factors influencing the physical properties of ferroic oxide nanostructures for strained SLs which retain superior physical properties compared to the parent materials. For instance, the effects of biaxial strains were found to tune the paraelectric-to-ferroelectric phase transition temperature ($T_C$), piezoelectricity, ferroelectricity, and dielectric properties of ferroelectric superlattices. Similarly, Cracium and co-workers [8] observed a variation in dielectric, ferroelectric and ferromagnetic properties as well as a change in paraelectric-ferroelectric temperature transition by introducing dopant and varying the unit cell parameters in perovskite ceramics. Besides, experimental and numerical investigations [9,10] showed that relaxed epitaxial ferroelectric films could exhibit a dielectric anomaly due to the presence of the inherent passivation layer in the heterostructure. In the SLs with few hundred nanometers thickness, the $T_C$ can be shifted by hundreds of degrees compared to bulk counterpart due to the strong polarization-strain coupling, suitable in practical applications [6,11]. The enhancement of the polarization via strain engineering was also predicted theoretically in ferroelectric SLs [12–14] and confirmed experimentally [15,16]. Earlier, in semiconductors the strain engineering had been intensively studied and had become an important parameter to enhance the performance of metal-oxide-semiconductor field-effect transistors (MOSFETs) for example via increasing the carriers' mobility of the strained Si, Ge or SiGe and optical properties in SiGe and in GaN [17–20].

In oxide thin films, generally the epitaxial strain can be engineered using different types of substrates having different thermal expansion and/or different lattice parameters but also defects and growth parameters such as temperature and oxygen pressure [9,21]. Moreover, in SLs there is an additional epitaxial strain originating from the lattice mismatch between the alternating layers, this strain can be tuned by varying the modulation period and/or the layer/layer thickness ratio. Several techniques were employed to investigate and determine qualitatively the strain in thin films such are XRD, neutron diffraction and Raman spectroscopy [11,22–25], because a precise distribution of this strain remains the key element to comprehend their effects on functional properties of the SLs materials to eventually tune their performances. Thus, for epitaxial ferroelectric nanomaterials, the use of transmission electron microscopy (TEM) techniques to probe the structural characteristic including the strains at atomic scale is of an essential interest to explain the ferroelectric and piezoelectric responses of the studied systems. Indeed, TEM was used for imaging the layers and interfaces in both thin films and SLs to reveal the presence of dislocations and strain relaxations for some critical thicknesses and determine the origin of ferroelectricity at nano-scale [2,4,26]. Moreover, TEM once combined to analytical tools allows measuring the strains generated in thin films at the atomic scale [27]. Among these techniques, high resolution transmission and scanning electron microscopy (HRTEM & HRSTEM) [28,29], convergent beam electron diffraction (CBED) [30], nano-beam electron diffraction (NBED) [31,32], dark-field electron holography (DFEH) [29], STEM Moiré interference [33–35] and geometrical phase analysis (HRTEM-GPA) [18,36,37]. In contrast to semiconductors, few recent works have been reported on strain measurements using TEM based techniques in ferroic materials indicating that this field still remains an experimental challenge [3,38].

The strain measurements performed using NBED technique is based on the analysis of acquired diffraction patterns collected from strained and unstrained (reference) areas of the sample. The accuracy and the sensitivity of the strain measurement using this technique were explored extensively by Williamson and co-workers [39]. These authors applied a diffraction reflection fitting algorithm to both simulated and experimental diffraction patterns collected from known composition of strained SiGe. These authors processed the recorded diffraction patterns with and without autocorrelation method. Their results showed that their approach allows achieving a strain sensitivity 4 times better with autocorrelation, which is greater than 0.06 % as cited in the literature and the average strain seems to be 10 to 40 times smaller with autocorrelation which are closer to their expected values of the strains for their sample under study. Interestingly, the comparison made by Favia and co-workers [40] regarding various methods used to measure the strains in thin films, showed that the conventional NBED technique remains one of the most sensitive techniques in strain measurements 7.5 $10^{-4}$. Only DFEH technique can do better 2.5 $10^{-4}$, but without the possibility of performing the strain mapping and it also requires heavy tools such as electrostatic biprism & Lorentz lens. Besides, CBED can provide similar resolution as DFEH but the field of view is very reduced compared to NBED, whereas HRTEM-GPA allows achieving a sensitivity of $10^{-3}$ [18,29].

To achieve 5 nm spatial resolution, Favia and co-workers [40] used in their experiment a small aperture of 10μm which is also limiting the analysis window to achieving larger strain mapping. The best sensitivity obtained on strain measurements using NBED technique is attributed to Beche and co-workers [41] who demonstrated a dramatical improvement of the spatial resolution obtained for NBED analysis to reach 2.7 nm with larger aperture of 50 μm by using a combination of Modern microscope FEI Titan and dedicated software including the algorithm for diffraction fitting and autocorrelation. These authors obtained a strain sensitivity of 6 x $10^{-4}$ and larger field a view compared to the one obtained elsewhere [40].

In the present work, HRSTEM combined with NBED technique associated with dedicated software Epsilon tools™ initially developed by FEI for semiconductors technology, and a cutting edge FEI Titan Microscope aberration corrected, are employed to quantify and map the elastic strains in two nuances of ferroelectric $[BaZrO_3]_{xΛ}/[BaTiO_3]_{(1-x)Λ}$ ($BZ_{xΛ}/BT_{(1-x)Λ}$) SLs having periods of $Λ = 6.6$ nm and $Λ = 11$ nm respectively, grown by pulsed laser deposition (PLD) on $SrTiO_3$ (STO) substrate to investigate the periodicity effect on the epitaxial strain and interfaces for a fixed total thickness of the SL and for BZ ratio of about x = 0.6. The dedicated software used in this study is similar to the one used by Beche and co-workers [41] and including the processing capabilities developed by Wialliamson and co-workers [39].

## 2. Experiment details

Two artificial $BZ_{x\Lambda}/BT_{(1-x)\Lambda}$ SLs of about 75 nm thickness were grown on (001) STO single crystal substrate by PLD technique using an excimer laser ($\lambda$= 248 nm) in a MECA 2000 UHV chamber. The modulation period ($\Lambda$) was fixed at 6.6 nm for the first SL (labelled BTBZ03) and at 11 nm for the second one (labelled BTBZ09) with x = 0.64 and 0.60, respectively. The alternating BT and BZ layers were deposited at the temperature of 750°C under oxygen pressure of 0.1 mbar. Reflection high-energy electron diffraction (RHEED) was used to control the surface quality of the layers at the end of the growth. High-resolution 4-circles diffractometer Bruker™ Discover D8 with a Cu $K_\alpha \lambda$ radiation was used for structural characterizations of the epitaxial SLs (reflectivity, $\omega/2\theta$, rocking curve and reciprocal space maps (RSMs). The out-of-plane lattice parameters of BT and BZ in SLs were obtained by modeling $\omega/2\theta$ X-ray patterns while the in-plane lattice parameters were determined from the RSMs measurements. The post mortem observations and the strains mapping of the sample's cross-section were performed using FEI™ Titan TEM microscope equipped with aberration correction and operating at 300 kV. Thin lamellae were prepared for both SLs using Dual Beam FEI™ Helios focus ion beam (FIB) combined with scanning electron microscope (SEM), following a conventional sample preparation method as described elsewhere[42]. The strains quantification and mapping were extracted during HRSTEM using NBED and processed off-line using dedicated software (Epsilon FEI™) including the processing capabilities as detailed in the introduction section. Strain data is acquired using the shifts in diffraction reflections as function of positions on the sample. The post-processing of the acquired diffraction patterns allows calculating the distance between diffraction spots to be compared to a reference diffraction pattern. As mentioned by Beche and co-workers [41], the reference diffraction pattern does not need necessarily to be collected from the material under investigation. The strain is then extracted and plotted as function of the (x,y) positions. In the sake of accuracy, the strain measurements error was determined and calibrated using standard calibration sample.

## 3. Results and discussion

### 3.1. X-ray analyses

Figure 1(a) shows the high-resolution x-ray reflectivity (XRR) diagram of BTBZ09 SL. The presence of clear finite thickness oscillations reveals that the smooth surfaces and interfaces are of a good quality. Satellite peaks are also visible indicating the chemical modulated structure along the growth direction.

The periodicity Λ of SLs can be determined using the angular distance between the satellite peaks based on the following formula:

$$\theta^2 = \left(\frac{\lambda}{2\Lambda}\right) \cdot n^2 + \theta_c^2 \qquad (1)$$

Where θ is the angular position of the satellite peaks, λ is the x-ray wavelength, $n$ the diffracted satellite peaks order, and $\theta_c$ is the critical angle. The insert of Figure 1(a) shows an example of the linear dependence of the $\theta^2$ as a function of $n^2$. The obtained periods from the linear fits are 6.60 nm and 11.19 nm for BTBZ03 and BTBZ09, respectively. The density, the thickness, and the roughness of BT and BZ layers can be estimated from the simulation of XRR. The obtained values are given in Table 1.

Figure 1(b) displays the room-temperature θ–2θ XRD patterns for BTBZ03 and BTBZ09 SLs with a thickness of about 72 nm and 77 nm, respectively grown on STO substrate. Both SLs revealed a single-phase with the presence of so-called satellite peaks; the evidence of modulated structures along the growth direction. The zoom-in around the satellite peaks shows the presence of the Laue oscillations (finite thickness oscillations) around the satellite peaks which is a signature of very good quality of the superlattice layers. The rocking curve performed on (002) satellite peaks for both SLs consists of two peaks as shown in the insert of Figure 1 (b). The first peak is narrow and has a full width at half maximum (FWHM) very close to the one obtained for the STO substrate. This peak corresponds to the strained epitaxial BT and BZ layers which present a very low mosaicity. The presence of a second broader peak seems to indicate the presence of relaxed layers at BT/BZ interfaces and the presence of dislocations at the substrate-SL interface as demonstrated in Figure S3 (supplementary information). In this Figure, both nuances of SLs were screened using HRSTEM technique combined with edge dislocations identification and localization tool as described elsewhere[43,44]. By combining HRTEM and different filters using FFT patterns, specific atomic arrangement directions are generated in form of discontinuities along atomic columns represented by continued lines. Matlab™ program was then used to streamline the counting and the positioning of dislocations. For both samples, the dislocations are present and their density is higher for the BTBZ09 sample. This is coherent with the stabilization of the width of the inter-diffusion interfaces and the strain relaxation measured (discussed in the following sections) as also observed in fully relaxed Ge film in SiGe heterostructure [45]. For BTBZ03 superlattice ($\Lambda$ = 6.6 nm) the full width at half maximum (FWHM) of the broader peak is larger than the one of BTBZ09 ($\Lambda$ = 11 nm). This is attributed to the number of interfaces, which is higher for the superlattice with $\Lambda$ = 6.6nm.

The out-of-plane lattice parameters of BT ($d_{BT}$) and BZ ($d_{BZ}$) layers in SLs are obtained from the simulation of XRD patterns using a model calculation developed elsewhere[24,46,47]. Figures 2(a) and (c) illustrate, respectively, the comparison between experimental and calculated first order diffractograms for both BTBZ03 and BTBZ09 SLs that are in good agreement. The obtained parameters from the model calculation, namely $d_{BT}$, $d_{BZ}$, in-plane lattice parameter $a_{SL}$, $\Lambda$ and layer thicknesses $e_{BT}$ and $e_{BZ}$, are given in Table 2. The layer thicknesses are obtained using the following equations

$$e_{BT} = N_{BT} \times d_{BT} \quad (2) \quad \text{and} \quad e_{BZ} = N_{BZ} \times d_{BZ} \quad (3)$$

where $N_{BT}$ and $N_{BZ}$ are the number of unit cell of BT and BZ layers, respectively.

For both SLs, the $d_{BT}$ is smaller than the *a*-axis of BT bulk while the $d_{BZ}$ is larger than BZ bulk lattice parameter. This implies that the BT layers are under extensive strain (c-axis lies in the plane of the substrate), while the BZ layers are compressed by the adjacent BT layers leading to distorted layers with c-axis orientation. These results are in agreement with our previous XRD and Raman investigations on BT/BZ SLs grown on MgO substrate [24,48]. Moreover, the in-plane lattice parameter of SLs is determined from XRD-RSM. Figures 2(b) and (d) display the RSM obtained around the (103) reflections for BTBZ03 and BTBZ09, respectively. For both SLs, the reflection of superlattice BT/BZ layers, originating from the satellite peaks diffraction, is clearly shifted along $q_x$ axis with respect to the STO substrate. This indicates a partial in-plane relaxation of the structure with respect to the STO substrate. This relaxation is due in part to the high lattice mismatch (7.3%) between the first layer (BZ) and the substrate (the strain obtained in our SLs is mainly induced by the lattice mismatch at interfaces of the alternating BT and BZ layers around 3.72%). Here, it is worth noticing that the superlattice reflections are broader along the $q_x$, thus the in-plane lattice parameter $a_{SL}$ is calculated using the middle of the reflections i.e. the intense zone. For the two SLs, $a_{SL}$ is found to be ~ 4.14Å. This value (e.g. Table 2), lies between the out-of-plane lattice parameter of BT and BZ layers in superlattice but it is larger and smaller than the values of c-axis of BT and BZ bulks respectively. This confirms that the BT and BZ layers are under in-plane extensive and compressive strain respectively.

### 3.2. Microstructure analyses

Figure 3 shows the microstructures as captured by HRTEM and HRSTEM for both SLs BTBZ03 (top) and BTBZ09 (bottom). The inserts (Figure3b and 3c) highlight the interface between the BT and BZ layers. It is clear from the HRTEM micrographs that these interfaces are free of the defects and irregularities, which indicates the high quality of the grown SLs. These micrographs confirm the morphology of the interfaces as revealed by XRR diagrams. It is worth noticing that the good quality of the interfaces BZ/BT is essential for the obtained SLs as they are one of the key parameters controlling their functional properties by being sites for inter-diffusion. Indeed, altering the crystal structure of the layers at the edges may influence for instance the paraelectric-to-ferroelectric phase transition temperature[6,47]. To examine the BT and BZ layers intermixing and the size of their interfaces for both samples, intensity profiles were extracted from HRSTEM micrographs. Figure 4 illustrates the HRSTEM images used Figures 4 (a) and 4 (c) to record the corresponding intensity profiles plotted in Figures 4 (b) and 4 (d) for BTBZ03 and BTBZ09 respectively. $\delta_i$ represents the width of the BT/BZ interfaces where i represents its order. Our findings shown in Figure 5, indicate that the BT/BZ interface width varies between 0.63 nm and 0.77 nm for the first 6 interfaces and it stabilizes at the 7th interface attaining an average value of 0.55 nm for BTBZ03, whereas for BTBZ09, the BT/BZ interface width oscillates between 0.95 nm and 1.21 nm. This can be attributed to relaxation mechanism: For high $\Lambda$ = 11 nm, the BT and BZ layers are larger enough to accommodate the strain and stabilize, leading to regular interfaces width where for low $\Lambda$ = 6.6 nm, the stability occurs only at long range (7[th] BT/BZ interface) due to the confinement of the layers. Indeed, the larger is the inter-diffusion interface with respect to the size of considered layer, the more strained adjacent layers are encountered. In other words, the larger interface allows higher atomic exchange between BT and BZ layers leading to the internal stresses relaxation and this could happen if the layer thickness is large enough to allow this diffusion process. For narrower layers, the mean free path of the atomic mobility is cut off by the presence of the next strained layer resulting in continuous variation of the strain whereas this mean free path is large enough to allow this relaxation. This statement is coherent with the stabilization of strains measured in BTBZ09 compared to the strain measured in BTBZ03 (next section).

### 3.3. Strain analyses

High-angle annular dark field (HAADF) STEM technique was used to image the cross section of BTBZ03 and BTBZ09 samples as illustrated in Figures 6 (a) and 6 (c) respectively. This technique is atomic number Z sensitive allowing to segregate materials as a function of their atomic number. As Zr is the heavier atom compared to Ti, hence the image contrast intensity due to the electrons scattering is higher for Zr, resulting in brighter contrast for BZ layers. Energy filtered transmission electron microscopy (EFTEM) was then applied to perform the elemental mapping and confirm the composition and the stacking sequence of the different layers for BTBZ03 and BTBZ09 shown in Figures 6 (b) and 6 (d) respectively. From these maps highlighting the BT and BZ layers, periods of $\Lambda = 6.6$ nm and $\Lambda = 11$ nm are measured for the samples BTBZ03 and BTBZ09, respectively, which is in accordance to the values expected by the processing and confirmed by XRD analyses.

Strain mapping are rendered from the variation of lattice parameters of different layers compared to STO substrate (reference) using the microscope in NBED mode as described above. The diffraction patterns were collected using 20 μm objective aperture with a semi convergence angle of 0.43 mrad. A FWHM beam size of 2 nm was measured (e.g. Figure S4 in supplementary information). Diffraction patterns are recoded using a 2k x 2k CCD camera mounted in the TEM. An area of 45 nm x 45 nm was analyzed with a step size of 2 nm. In total, 484 diffraction patterns for each set of samples were acquired and analyzed using the integrated algorithm as described above. Figure 7 gives the strain maps recorded and rendered in color codes for BTBZ03 and BTBZ09 with in-plane (horizontal) and out-of-plane (vertical) strain plotted as a function of the position. Figures 7(a) and 7

(c) show a linear in-plane deformation of 5.5% and 6.1% for BTBZ03 and BTBZ09, respectively. This in agreement with the XRD reciprocal space mappings, which have shown a different in-plane lattice parameters of the SLs compared to the STO substrate. The out-of-plane deformation maps plotted in Figures7 (b) and 7 (d) show a nonlinear behavior directly linked to the composition of the layers, where BT and BZ layers are clearly identifiable. The layers thicknesses and periods are in accordance with the values measured on HAADF-STEM images. For sample BTBZ03, BZ layers have a strain ranging from 5.7% to 6.8% and BT layers have a strain ranging from 4.6% to 5.4%. The obtained strain for BT layers in BTBZ03 is quite high because their size lies at the limit of the resolution of the strain measurements. This is also confirmed by the strain mapping plotted for BTBZ03 in Figure 7 (b) where one can barely distinguish the strains for the BT layers. In contrast, for sample BTBZ09 both layers BZ and BT have larger size showing fairly strain variation, namely, BZ layers have a strain ranging from 6.3% to 7% and BT layers have a strain ranging from 1.7% to 4.8%.

The layers thicknesses and period obviously influence the deformation of the different layers. To further investigate these effects, lattice parameters have been extracted/computed from Figure 7 and plotted in Figure 8. The data are integrated in the vertical direction to obtain more representative lattice parameter values. The lattice parameter of STO obtained using XRD ($a_{STO}$ = 3.905 Å) was used to retrieve BT and BZ lattice parameter variations. For comparison purposes, lattice parameters of BT and BZ obtained from the literature and XRD measurements are also plotted. In-plane lattice parameters are in close accordance with XRD measurements for both BTBZ03 and BTBZ09. The measured values of 4.14 Å for BTBZ03 and 4.11 Å for BTBZ09 are in between $a_{BZ,bulk}$ and $a_{BT,bulk}$ [24], which indicates that BZ layers are compressively strained whereas BT layers are tensile strained. The two slopes of the curves show a slight decrease after the first set of BT/BZ layers but this drop is more pronounced for BTBZ09, which indicates a higher relaxation for the following BT layers. The out-of-plane lattice parameters exhibit a different behavior: crystal lattices have the freedom to change in the vertical direction as the substrate doesn't vertically constrain the SLs. Thus, the out-of-plane lattice parameter curves show peak/valley duos corresponding to BT/BZ number of layers. The out-of-plane lattice parameter of BZ shows a similar trend for BTBZ03 and BTBZ09. The first grown BZ layers have a lattice parameter ~ 4.14 Å. This parameter increases while approaching the surface to reach a maximum of 4.17 Å. For both cases, $c_{BZ}$ is lower than the vertical lattice parameter given in the literature as determined by XRD technique, but higher than the in-plane lattice parameters which indicates a vertical compressive deformation of BZ layers for both samples. The difference between the two samples is observable for the out-of-plane lattice parameter of BT layers. Indeed, the $c_{BT03}$ varies from 4.08 Å to 4.12 Å whereas $c_{BT09}$ varies from 3.97 Å to 4.09 Å. It is worth noticing that the determined out-plane lattice parameter (as absolute value) for BT layers in BTBZ03 SLs (≈ 2.2 nm thick) needs to be considered with caution considering both the resolution of the technique used in this study (≈ 2 nm) and the strain measurement errors (0.15 %) determined on the magical sample. Hence, the analysis of the strain results obtained with NBED technique at 2 nm resolution demonstrated that the strain measurements as well as the strain mapping in complex SLs can be resolved with confidence for BZ layers for both SLs and for BT layers in BTBZ09 (size less than 4 nm) but shows some limitations for the BT layer in BTBZ03 (size less than 2.2 nm).

Besides, the variations in the lattice parameters along the growth direction observed in both BTBZ03 and BTBZ09 SLs, could be triggered by the presence of domain structures within the ferroelectric layers [49,50]. Indeed, likewise thin film ferroelectric layers, SLs may accommodate strain through ferroelastic domains as demonstrated elsewhere [3,6,49]. The HRTEM micrographs given in Figure 3b, seem to exhibit some vertical striations that could be associated with the strain modulations along the in-plane direction suggesting a possible presence of domain structures.

These results are strongly linked to the interfaces size for both samples. Indeed, the average inter-diffusion interfaces was found to cover 8 to 12% and 9 to 11% of the total thickness of the BTBZ03 and BTBZ09 samples respectively, which could affect significantly the physical properties of the material as reported in our earlier investigation, where the dielectric permittivity was found to be strongly affected by the SL period[51]. This could be explained by the presence of a large number of inter-diffusion interfaces due to the small period as the capacitance of the entire SL material is a series of capacitances of BT layer, BZ layer and BT/BZ interfaces.

## 4. Conclusion

A local survey at atomic scale of the elastic strains in ferroelectric SLs was demonstrated using HRSTEM and NBED in conjunction with integrated software showing that strain is strongly correlated to the SL periods. Recall that a precise distribution of the strain in ferroic oxide materials is one of the key parameters that could contribute to tuning the functional properties of SLs. This work allowed capturing the elastic strains both in-plane and out-of-plane to be correlated to the inter-diffusion interfaces. The choice of thin film SL was motivated by their complex structure composed of alternating layers of two compounds in which the elastic strain is presumably different. In addition, compared to single films, SLs systems offer the possibility to study the strain at both the substrate-layer and layer-layer interfaces in which elastic deformations are quite different from their respective bulk materials. Earlier, we showed that in the BT/BZ SLs the strain and interfaces are important parameters to tune the structural properties of SLs such as vibrational and electrical properties since the BT and BZ materials present significant lattice mismatch (BT bulk is tetragonal, with $a_{BT}$ = 0.3992 nm and $c_{BT}$ = 0.4036 nm and BZ bulk is cubic, with $a_{BZ}$ = 0.4192 nm). For instance, the dielectric permittivity, ferroelectric polarization and energy storage properties of these systems have been correlated to the competition between the effect of the strain and numbers of interfaces controlled by the variation of the modulation period for a fixed total thickness of the sample.

**Acknowledgements**

This work was supported by European Union's H2020-MSCA-RISE (ENIGMA No 778072).

# References

[1] A.R. Damodaran, J.D. Clarkson, Z. Hong, H. Liu, A.K. Yadav, C.T. Nelson, S.-L. Hsu, M.R. McCarter, K.-D. Park, V. Kravtsov, A. Farhan, Y. Dong, Z. Cai, H. Zhou, P. Aguado-Puente, P. García-Fernández, J. Íñiguez, J. Junquera, A. Scholl, M.B. Raschke, L.-Q. Chen, D.D. Fong, R. Ramesh, L.W. Martin, Phase coexistence and electric-field control of toroidal order in oxide superlattices, Nat. Mater. 16 (2017) 1003. https://doi.org/10.1038/nmat4951.

[2] G.Y. Kim, K. Chu, K.D. Sung, H.S. Lee, S.D. Kim, K. Song, T. Choi, J. Lee, J.P. Buban, S.Y. Yoon, K.H. Kim, C.H. Yang, S.Y. Choi, Disordered ferroelectricity in the PbTiO3/SrTiO3 superlattice thin film, APL Mater. 5 (2017). doi:10.1063/1.4986064.

[3] Q. Li, C.T. Nelson, S.L. Hsu, A.R. Damodaran, L.L. Li, A.K. Yadav, M. McCarter, L.W. Martin, R. Ramesh, S. V. Kalinin, Quantification of flexoelectricity in PbTiO3/SrTiO3 superlattice polar vortices using machine learning and phase-field modeling, Nat. Commun. 8 (2017). doi:10.1038/s41467-017-01733-8.

[4] A.K. Yadav, C.T. Nelson, S.L. Hsu, Z. Hong, J.D. Clarkson, C.M. Schlepütz, A.R. Damodaran, P. Shafer, E. Arenholz, L.R. Dedon, D. Chen, A. Vishwanath, A.M. Minor, L.Q. Chen, J.F. Scott, L.W. Martin, R. Ramesh, Observation of polar vortices in oxide superlattices, Nature. 530 (2016) 198. https://doi.org/10.1038/nature16463.

[5] A.J. Grutter, H. Yang, B.J. Kirby, M.R. Fitzsimmons, J.A. Aguiar, N.D. Browning, C.A. Jenkins, E. Arenholz, V. V Mehta, U.S. Alaan, Y. Suzuki, Interfacial Ferromagnetism in ${\mathrm{LaNiO}}_{3}/{\mathrm{CaMnO}}_{3}$ Superlattices, Phys. Rev. Lett. 111 (2013) 87202. doi:10.1103/PhysRevLett.111.087202.

[6] J. Belhadi, M. El Marssi, Y. Gagou, Y.I. Yuzyuk, I.P. Raevski, Giant increase of ferroelectric phase transition temperature in highly strained ferroelectric [BaTiO 3 ] 0.7Λ /[BaZrO 3 ] 0.3Λ superlattice, Europhys. Lett. 106 (2014) 17004. http://stacks.iop.org/0295-5075/106/i=1/a=17004.

[7] R. Oja, M. Tyunina, L. Yao, T. Pinomaa, T. Kocourek, A. Dejneka, O. Stupakov, M. Jelinek, V. Trepakov, S. Van Dijken, R.M. Nieminen, D0 ferromagnetic interface between nonmagnetic perovskites, Phys. Rev. Lett. 109 (2012) 1–5. doi:10.1103/PhysRevLett.109.127207.

[8] F. Craciun, M. Cernea, V. Fruth, M. Zaharescu, I. Atkinson, N. Stanica, L.C. Tanase, L. Diamandescu, A. Iuga, C. Galassi, Novel multiferroic (Pb1−3x/2Ndx)(Ti0.98−yFeyMn0.02)O3 ceramics with coexisting ferroelectricity and ferromagnetism at ambient temperature, Mater. Des. 110 (2016) 693–704. doi:10.1016/j.matdes.2016.08.046.

[9] Y.I. Y. Gagou, J. Belhadi, B. Asbani, M. El Marssi, J-L Dellis, J.. S. Yuzyuk, I.P. Raevski, Intrinsic dead layer effects in relaxed epitaxial BaTiO3 thin film grown by pulsed laser deposition, Mater. Des. 122 (2017) 157–163. doi:10.1016/j.matdes.2017.03.001.

[10] D. Xiao, L. Meng, G. Yu, The effect of surface layer on the dielectric behavior of complex oxide thin films, Mater. Des. 24 (2003) 377-382m. doi:10.1016/S0261-3069(03)00030-X.

[11] D.A. Tenne, X. Xi, Raman Spectroscopy of Ferroelectric Thin Films and Superlattices, J. Am. Ceram. Soc. 91 (2008) 1820–1834. doi:10.1111/j.1551-2916.2008.02371.x.

[12] J.B. Neaton, K.M. Rabe, Theory of polarization enhancement in epitaxial BaTiO3/SrTiO3 superlattices, Appl. Phys. Lett. 82

(2003) 1586–1588. doi:10.1063/1.1559651.

[13] S.M. Nakhmanson, K.M. Rabe, D. Vanderbilt, Polarization enhancement in two- and three-component ferroelectric superlattices, Appl. Phys. Lett. 87 (2005) 102906. doi:10.1063/1.2042630.

[14] S.M. Nakhmanson, K.M. Rabe, D. Vanderbilt, Predicting polarization enhancement in multicomponent ferroelectric superlattices, Phys. Rev. B. 73 (2006) 60101. doi:10.1103/PhysRevB.73.060101.

[15] T. Shimuta, O. Nakagawara, T. Makino, S. Arai, H. Tabata, T. Kawai, Enhancement of remanent polarization in epitaxial BaTiO3/SrTiO3 superlattices with "asymmetric" structure, J. Appl. Phys. 91 (2002) 2290–2294. doi:10.1063/1.1434547.

[16] H.N. Lee, H.M. Christen, M.F. Chisholm, C.M. Rouleau, D.H. Lowndes, Strong polarization enhancement in asymmetric three-component ferroelectric superlattices, Nature. 433 (2005) 395. https://doi.org/10.1038/nature03261.

[17] M.L. Lee, E.A. Fitzgerald, M.T. Bulsara, M.T. Currie, A. Lochtefeld, Strained Si, SiGe, and Ge channels for high-mobility metal-oxide-semiconductor field-effect transistors, J. Appl. Phys. 97 (2004) 11101. doi:10.1063/1.1819976.

[18] F. Hüe, M. Hÿtch, H. Bender, F. Houdellier, A. Claverie, Direct Mapping of Strain in a Strained Silicon Transistor by High-Resolution Electron Microscopy, Phys. Rev. Lett. 100 (2008) 156602. doi:10.1103/PhysRevLett.100.156602.

[19] F. Ravaux, N.S. Rajput, J. Abed, L. George, M. Tiner, M. Jouiad, Effect of rapid thermal annealing on crystallization and stress relaxation of SiGe nanoparticles deposited by ICP PECVD, RSC Adv. 7 (2017) 32087–32092. doi:10.1039/c7ra04426g.

[20] A. Najar, M. Gerland, M. Jouiad, Porosity-induced relaxation of strains in GaN layers studied by means of micro-indentation and optical spectroscopy, J. Appl. Phys. 111 (2012). doi:10.1063/1.4710994.

[21] B. Allouche, Y. Gagou, F. Le Marrec, M. Fremy, M. El Marssi, Bipolar resistive switching and substrate effect in GdK 2 Nb 5 O 15 epitaxial thin fi lms with tetragonal tungsten bronze type structure, Mater. Des. 112 (2016) 80–87. doi:10.1016/j.matdes.2016.09.047.

[22] B. Fluegel, A. V Mialitsin, D.A. Beaton, J.L. Reno, A. Mascarenhas, Electronic Raman scattering as an ultra-sensitive probe of strain effects in semiconductors, Nat. Commun. 6 (2015) 7136. https://doi.org/10.1038/ncomms8136.

[23] Y.I. Yuzyuk, R.A. Sakhovoy, O.A. Maslova, V.B. Shirokov, I.N. Zakharchenko, J. Belhadi, M. El Marssi, Phase transitions in BaTiO3 thin films and BaTiO3/BaZrO3 superlattices, J. Appl. Phys. 116 (2014) 184102. doi:10.1063/1.4901207.

[24] M. El Marssi, Y. Gagou, J. Belhadi, F. De Guerville, Y.I. Yuzyuk, I.P. Raevski, Ferroelectric BaTiO3/BaZrO3 superlattices: X-ray diffraction, Raman spectroscopy, and polarization hysteresis loops, J. Appl. Phys. 108 (2010) 84104. doi:10.1063/1.3496620.

[25] R.R. Das, Y.I. Yuzyuk, P. Bhattacharya, V. Gupta, R.S. Katiyar, Folded acoustic phonons and soft mode dynamics in BaTiO 3/SrTiO3 superlattices, Phys. Rev. B - Condens. Matter Mater. Phys. 69 (2004) 6–9. doi:10.1103/PhysRevB.69.132302.

[26] A. Herklotz, D. Lee, E.-J. Guo, T.L. Meyer, J.R. Petrie, H.N. Lee, Strain coupling of oxygen non-stoichiometry in perovskite thin films, J. Phys. Condens. Matter. 29 (2017) 493001. http://stacks.iop.org/0953-8984/29/i=49/a=493001.

[27] V.B. Ozdol, C. Gammer, X.G. Jin, P. Ercius, C. Ophus, J. Ciston, A.M. Minor, Strain mapping at nanometer resolution using advanced nano-beam electron diffraction, Appl. Phys. Lett. 106 (2015). doi:10.1063/1.4922994.

[28] R. Bierwolf, M. Hohenstein, F. Phillipp, O. Brandt, G.E. Crook, K. Ploog, Direct measurement of local lattice distortions in

strained layer structures by HREM, Ultramicroscopy. 49 (1993) 273–285. doi:https://doi.org/10.1016/0304-3991(93)90234-O.

[29] M. Hÿtch, F. Houdellier, F. Hüe, E. Snoeck, Nanoscale holographic interferometry for strain measurements in electronic devices, Nature. 453 (2008) 1086. https://doi.org/10.1038/nature07049.

[30] A. Armigliato, R. Balboni, G.P. Carnevale, G. Pavia, D. Piccolo, S. Frabboni, A. Benedetti, A.G. Cullis, Application of convergent beam electron diffraction to two-dimensional strain mapping in silicon devices, Appl. Phys. Lett. 82 (2003) 2172–2174. doi:10.1063/1.1565181.

[31] K. Usuda, T. Numata, T. Irisawa, N. Hirashita, S. Takagi, Strain characterization in SOI and strained-Si on SGOI MOSFET channel using nano-beam electron diffraction (NBD), Mater. Sci. Eng. B. 124–125 (2005) 143–147. doi:https://doi.org/10.1016/j.mseb.2005.08.062.

[32] F. Uesugi, A. Hokazono, S. Takeno, Evaluation of two-dimensional strain distribution by STEM/NBD, Ultramicroscopy. 111 (2011) 995–998. doi:https://doi.org/10.1016/j.ultramic.2011.01.035.

[33] N. Cherkashin, T. Denneulin, M.J. Hÿtch, Electron microscopy by specimen design: Application to strain measurements, Sci. Rep. 7 (2017) 1–8. doi:10.1038/s41598-017-12695-8.

[34] S. Kim, J.J. Kim, Y. Jung, K. Lee, G. Byun, K. Hwang, S. Lee, K. Lee, Reliable strain measurement in transistor arrays by robust scanning transmission electron microscopy, AIP Adv. 3 (2013) 92110. doi:10.1063/1.4821278.

[35] D. Su, Y. Zhu, Scanning moiré fringe imaging by scanning transmission electron microscopy, Ultramicroscopy. 110 (2010) 229–233. doi:10.1016/j.ultramic.2009.11.015.

[36] D. Cooper, A. Béché, J.M. Hartmann, V. Carron, J.L. Rouvìre, Strain evolution during the silicidation of nanometer-scale SiGe semiconductor devices studied by dark field electron holography, Appl. Phys. Lett. 96 (2010) 2010–2013. doi:10.1063/1.3358149.

[37] M.J. Hÿtch, J.-L. Putaux, J.-M. Pénisson, Measurement of the displacement field of dislocations to 0.03 Å by electron microscopy, Nature. 423 (2003) 270–273. doi:10.1038/nature01638.

[38] H.J. Lee, S.S. Lee, J.H. Kwak, Y.-M. Kim, H.Y. Jeong, A.Y. Borisevich, S.Y. Lee, D.Y. Noh, O. Kwon, Y. Kim, J.Y. Jo, Depth resolved lattice-charge coupling in epitaxial BiFeO3 thin film, Sci. Rep. 6 (2016) 38724. https://doi.org/10.1038/srep38724.

[39] P.V.D. M. J. Williamson, J. Flanagan, Quantitative Analysis of the Accuracy and Sensitivity of Strain Measurements from, Symp. Phys. Fail. Anal. Integr. Circuits, IEEE. (2015) 197–200.

[40] P. Favia, M.B. Gonzales, E. Simoen, P. Verheyen, D. Klenov, H. Bender, J.E. Soc, P. H-h, P. Favia, M.B. Gonzales, E. Simoen, P. Verheyen, D. Klenov, Nanobeam Diffraction : Technique Evaluation and Strain Measurement on Complementary Metal Oxide Semiconductor Devices, J. Electrochem. Soc. 158 (2011) 438–446. doi:10.1149/1.3546851.

[41] A. Béché, J.L. Rouvière, L. Clément, J.M. Hartmann, A. Béché, J.L. Rouvière, L. Clément, J.M. Hartmann, Improved precision in strain measurement using nanobeam electron diffraction, Appl. Phys. Lett. 123114 (2013) 20–23. doi:10.1063/1.3224886.

[42] N.S. Rajput, Y. Shao-Horn, X.H. Li, S.G. Kim, M. Jouiad, Investigation of plasmon resonance in metal/dielectric nanocavities for high-efficiency photocatalytic device, Phys. Chem. Chem. Phys. 19 (2017) 16989–16999. doi:10.1039/c7cp03212a.

[43] M. Jouiad, E. Marin, R.S. Devarapalli, J. Cormier, F. Ravaux, C. Le Gall, J.M. Franchet, Microstructure and mechanical properties

evolutions of alloy 718 during isothermal and thermal cycling over-aging, Mater. Des. 102 (2016) 284–296. doi:10.1016/j.matdes.2016.04.048.

[44] E. Snoeck, B. Warot, H. Ardhuin, A. Rocher, M.J. Casanove, R. Kilaas, M.J. Hÿtch, Quantitative analysis of strain field in thin films from HRTEM micrographs, Thin Solid Films. 319 (1998) 157–162. doi:https://doi.org/10.1016/S0040-6090(97)01113-9.

[45] Q. Liu, C. Zhao, S. Su, J. Li, Y. Xing, B. Cheng, Strain Field Mapping of Dislocations in a Ge / Si Heterostructure, PLoS One. 8 (2013) 1–6. doi:10.1371/journal.pone.0062672.

[46] F. De Guerville, M. El Marssi, I.P. Raevski, M.G. Karkut, Y.I. Yuzyuk, Soft mode dynamics and the reduction of ${\mathrm{Ti}}^{4+}$ disorder in ferroelectric/relaxor superlattices $\mathrm{Ba}\mathrm{Ti}{\mathrm{O}}_{3}/\mathrm{Ba}{\mathrm{Ti}}_{0.68}{\mathrm{Zr}}_{0.32}{\mathrm{O}}_{3}$, Phys. Rev. B. 74 (2006) 64107. doi:10.1103/PhysRevB.74.064107.

[47] I.K. Schuller, New Class of Layered Materials, Phys. Rev. Lett. 44 (1980) 1597–1600. doi:10.1103/PhysRevLett.44.1597.

[48] J. Belhadi, M. El Marssi, Y. Gagou, Y.I. Yuzyuk, Y. El Mendili, I.P. Raevski, H. Bouyanfif, J. Wolfman, Highly constrained ferroelectric [BaTiO3](1−x)Λ/[BaZrO3]xΛ superlattices: X-ray diffraction and Raman spectroscopy, J. Appl. Phys. 116 (2014) 34108. doi:10.1063/1.4890513.

[49] A. Torres-Pardo, A. Gloter, P. Zubko, N. Jecklin, C. Lichtensteiger, C. Colliex, J.-M. Triscone, O. Stéphan, Spectroscopic mapping of local structural distortions in ferroelectric PbTiO${}_{3}$/SrTiO${}_{3}$ superlattices at the unit-cell scale, Phys. Rev. B. 84 (2011) 220102. doi:10.1103/PhysRevB.84.220102.

[50] P. Aguado-Puente, J. Junquera, Structural and energetic properties of domains in PbTiO${}_{3}$/SrTiO${}_{3}$ superlattices from first principles, Phys. Rev. B. 85 (2012) 184105. doi:10.1103/PhysRevB.85.184105.

[51] M. Benyoussef, J. Belhadi, A. Lahmar, M. El Marssi, Tailoring the dielectric and energy storage properties in BaTiO3/BaZrO3superlattices, Mater. Lett. 234 (2019) 279–282. doi:10.1016/j.matlet.2018.09.123.

**Figure captions**

**Fig. 1**: (a) XRR pattern recorded for BTBZ09 SL, (b) θ–2θ x-ray diffraction (XRD) pattern for the BTBZ03 and BTBZ09 SLs. The insert in (a) displays the linear fit of the $\theta^2$ as a function of $n^2$. The insert in (b) displays rocking curves scan around (200) of STO substrate and the most intense satellite peaks of BTBZ03 and BTBZ09 SLs.

**Fig. 2**: First order θ-2θ x-ray diffraction patterns of BTBZ03 (a) and BTBZ09(c) SLs (black peaks) together with the results of the model calculations (red peaks). (b) and (d) show the Reciprocal Space Maps (RSM) around the (103) reflection for BTBZ03 and BTBZ09 SLs, respectively.

**Fig. 3**: a) HRTEM image showing the stacking layers BT/BZ, b) Zoom-in of the red square box showing the inter-diffusion interface between BT/BZ free of defects, c) HRSTEM image of the BT/BZ stack showing the high quality of grown SLs. BTBZ03 (left image) and BTBZ09 (right image).

**Fig. 4**: a) and c) HRSTEM images showing highlighting the interfaces BT/BZ, respectively for BTBZ03 and BTBZ09 samples, b) and d) Intensity profile extracted from HRSTEM micrographs, respectively for BTBZ03 and BTBZ09 samples.

**Fig.5**: Variation of the interface width as function of interface order for BTBZ03 (red) and BTBZ09 (blue).

**Fig. 6**: HAADF-STEM images of samples BTBZ03 (a) and BTBZ09 (b) showing the different layers of the superlattices. Images (c) and (d) show their respective EFTEM elemental mapping (Zr and Ti).

**Fig. 7**: In-plane and out-of-plane strain mappings for samples BTBZ03 (a) and (b) and BTBZ09(c) and (d), respectively. The deformation of the crystal structure is calculated using the substrate (STO) as a reference (dark blue region at the bottom of the images).

**Fig. 8**: Evolution of the in-plane and out-of-plane lattice parameters as a function of the position along the vertical axis of the samples for samples BTBZ03 (a) and (b) and BTBZ09 (c) and (d), respectively. These curves are calculated using STO lattice parameter determines using XRD data and strain maps obtained using NBED technique integrated along the vertical axis.

**Table captions**

**Table 1**: XRR parameters obtained for BaTiO$_3$ and BaZrO$_3$ layers from the simulations performed on the two SLs (figure SII).

**Table 2**: Lattices parameters, periodicity and thickness of layers of BTBZ03 and BTBZ09 SLs.

**Figures**

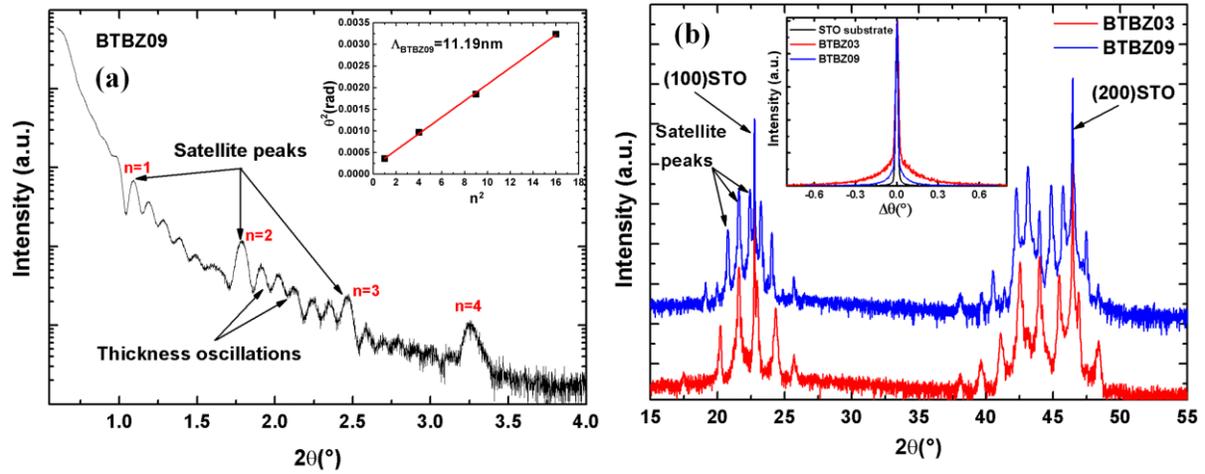

**Fig. 1**: (a) XRR pattern recorded for BTBZ09 SL, (b) θ–2θ x-ray diffraction (XRD) pattern for the BTBZ03 and BTBZ09 SLs. The insert in (a) displays the linear fit of the $\theta^2$ as a function of $n^2$. The insert in (b) displays rocking curves scan around (200) of STO substrate and the most intense satellite peaks of BTBZ03 and BTBZ09 SLs.

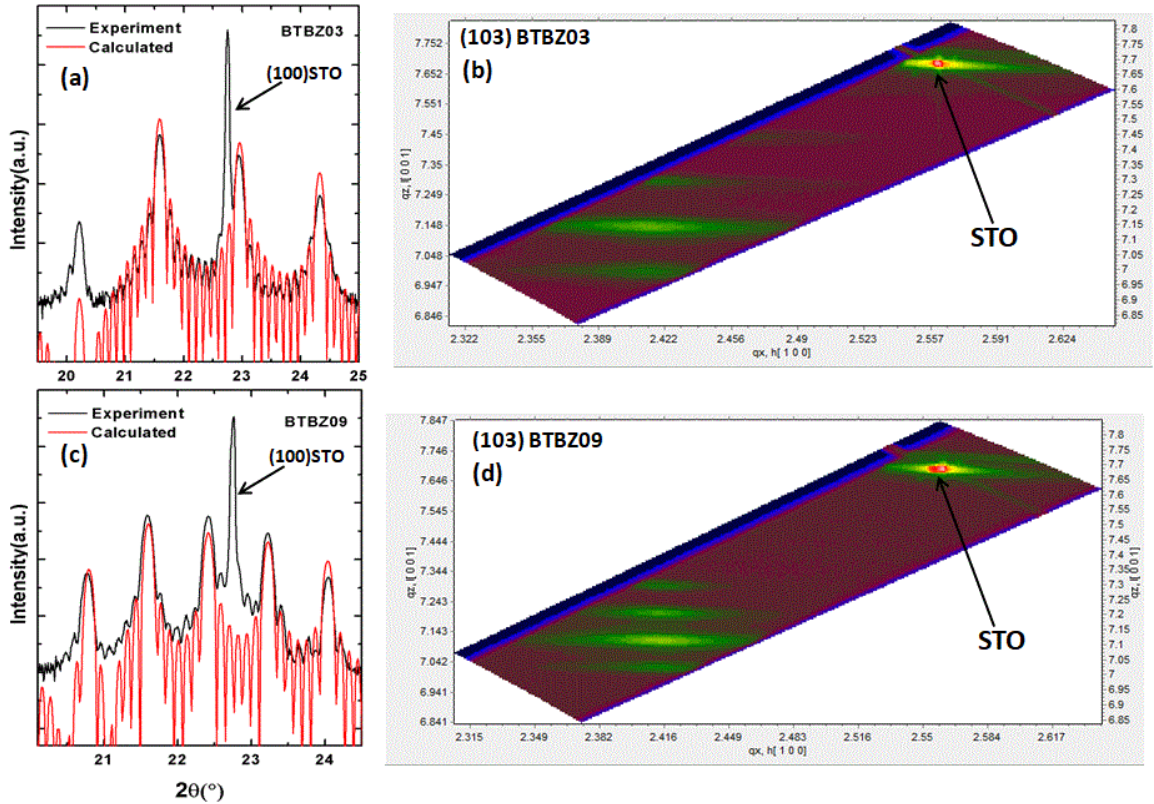

**Fig. 2**: First order θ-2θ x-ray diffraction patterns of BTBZ03 (a) and BTBZ09(c) SLs (black peaks) together with the results of the model calculations (red peaks). (b) and (d) show the Reciprocal Space Maps (RSM) around the (103) reflection for BTBZ03 and BTBZ09 SLs, respectively.

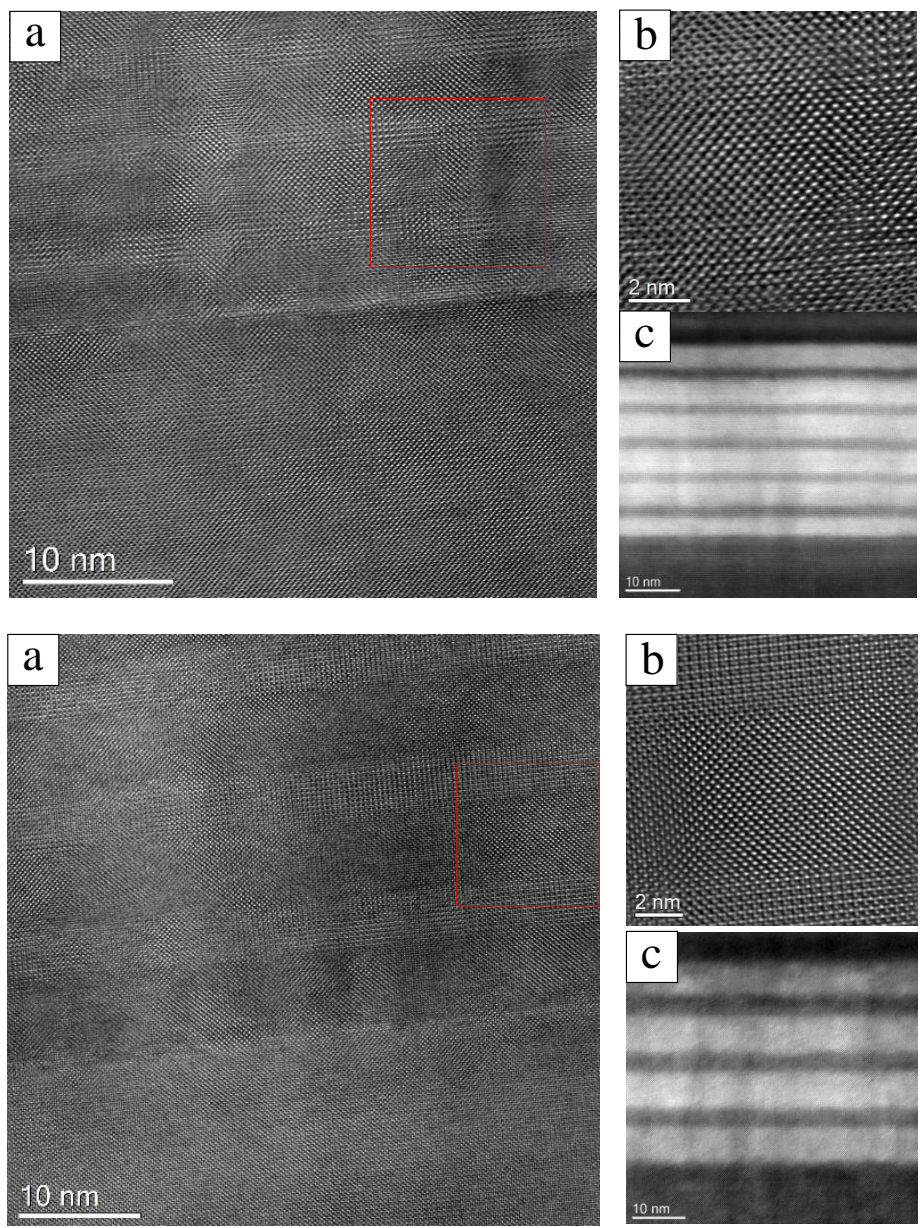

**Fig. 3**: a) HRTEM images showing the stacking layers BT/BZ, b) Zoom-in of the red square box showing the inter-diffusion interface between BT/BZ free of defects, c) HRSTEM image of the BT/BZ stack showing the high quality of grown SLs. BTBZ03 (Top image) and BTBZ09 (bottom image).

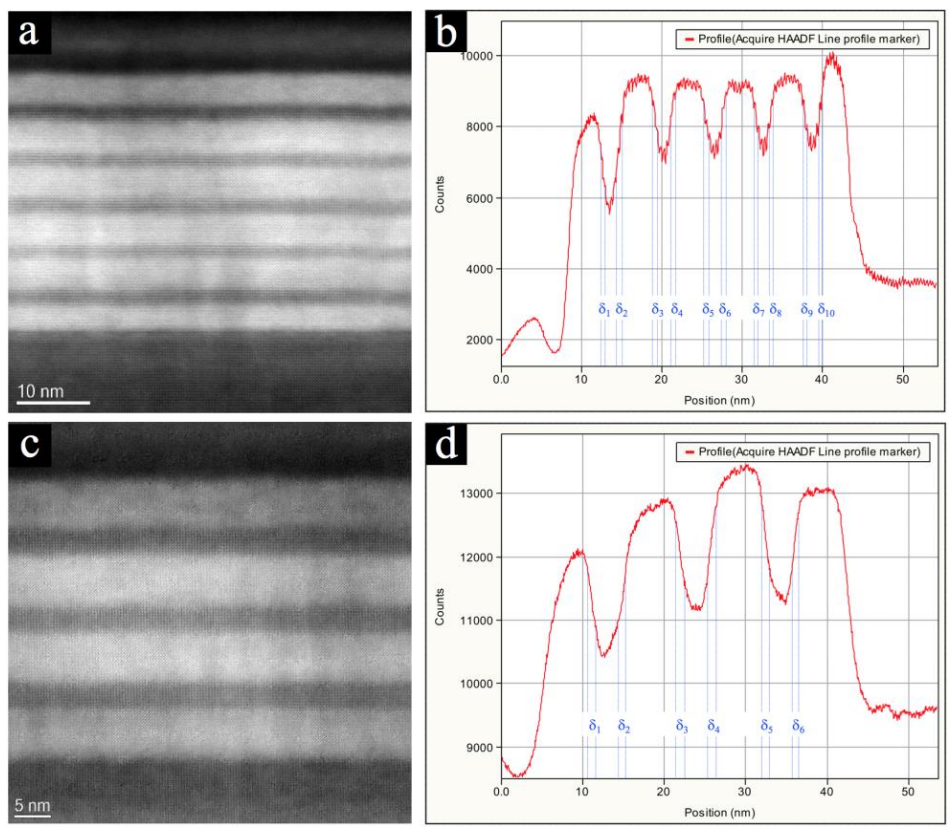

**Fig. 4**: a) and c) HRSTEM images showing highlighting the interfaces BT/BZ, respectively for BTBZ03 and BTBZ09 samples, b) and d) Intensity profile extracted from HRSTEM micrographs, respectively for BTBZ03 and BTBZ09 samples

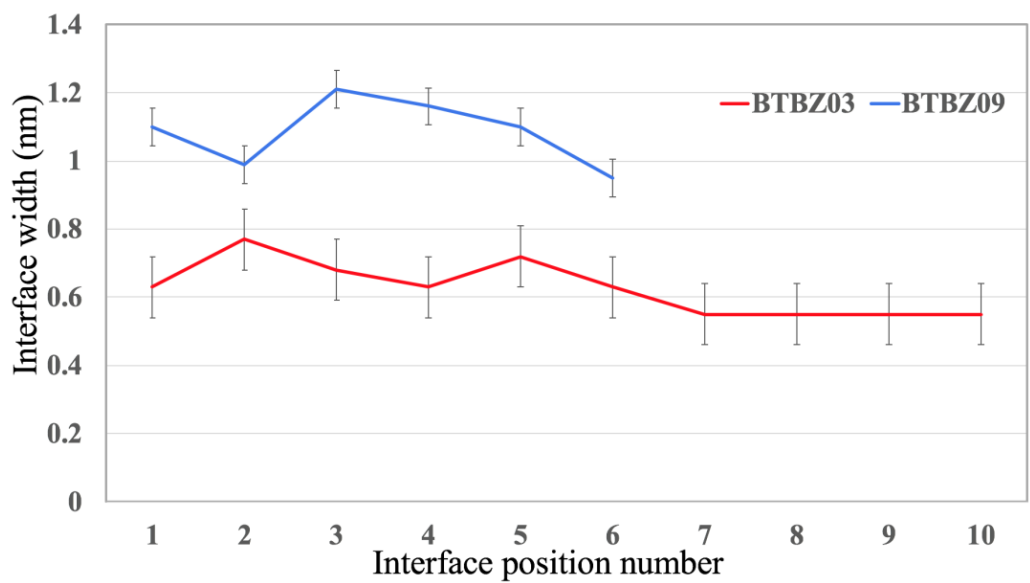

**Fig.5**: Variation of the interface width as function of interface order for BTBZ03 (red) and BTBZ09 (blue).

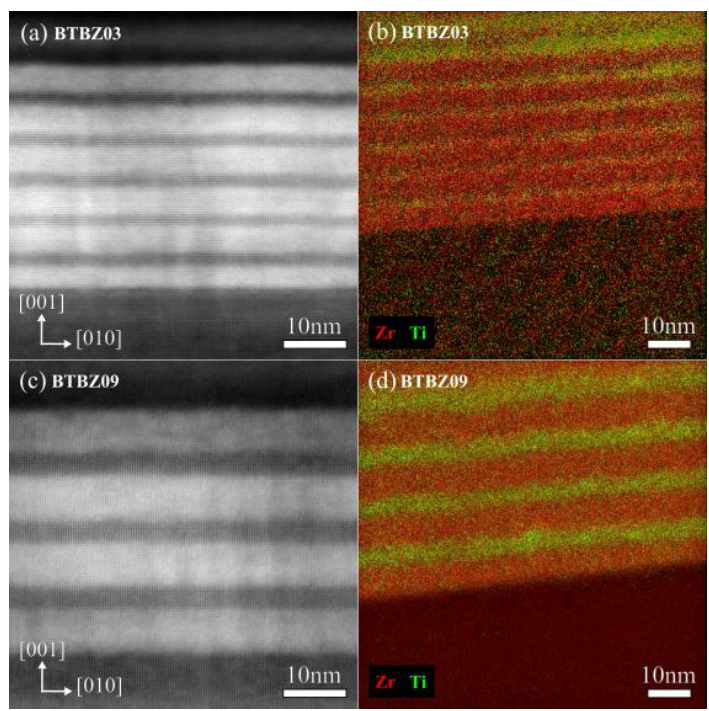

**Fig. 6**: HAADF-STEM images of samples BTBZ03 (a) and BTBZ09 (b) showing the different layers of the superlattices. Images (c) and (d) show their respective EFTEM elemental mapping (Zr and Ti).

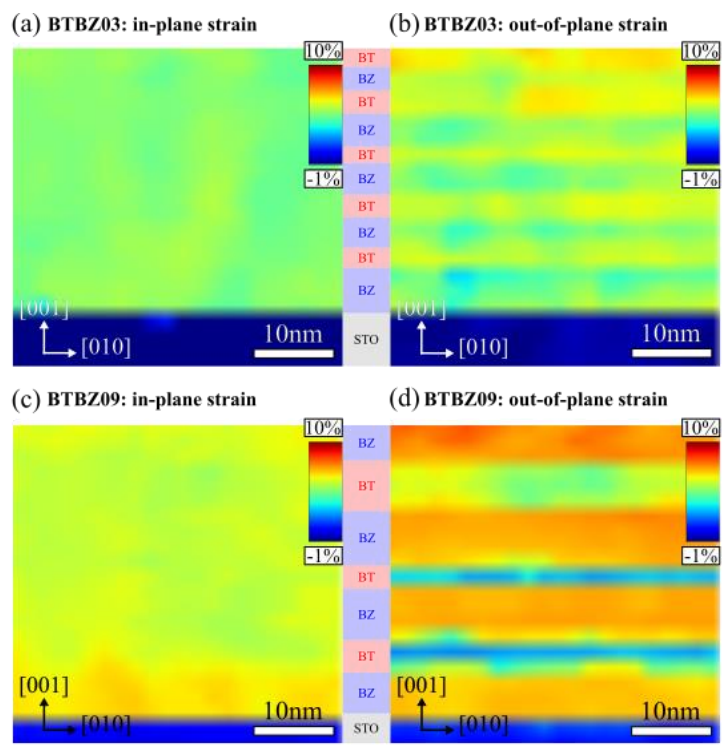

**Fig. 7**: In-plane and out-of-plane strain mappings for samples BTBZ03 (a) and (b) and BTBZ09(c) and (d), respectively. The deformation of the crystal structure is calculated using the substrate (STO) as a reference (dark blue region at the bottom of the images).

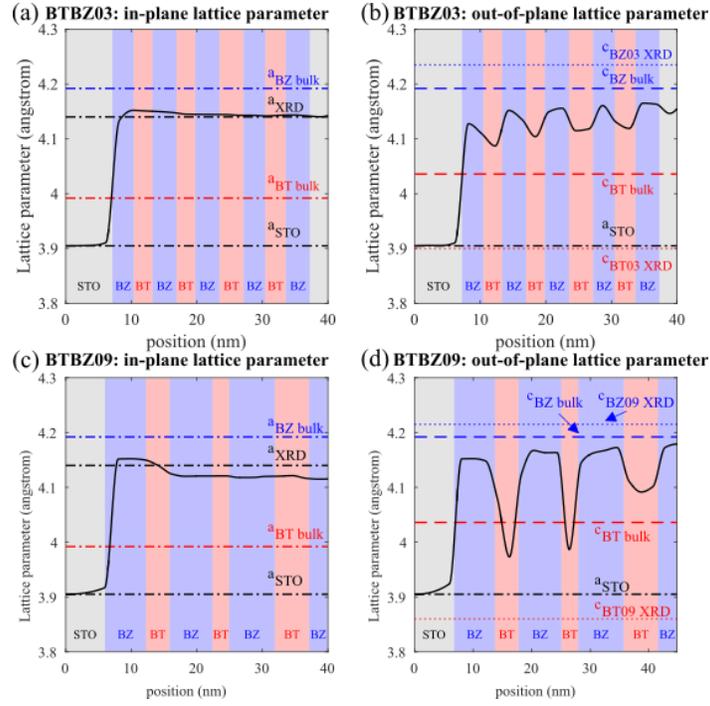

**Fig. 8**: Evolution of the in-plane and out-of-plane lattice parameters as a function of the position along the vertical axis of the samples for samples BTBZ03 (a) and (b) and BTBZ09 (c) and (d), respectively. These curves are calculated using STO lattice parameter determines using XRD data and strain maps obtained using NBED technique integrated along the vertical axis.

# Tables

Table 1: XRR parameters obtained for $BaTiO_3$ and $BaZrO_3$ layers from the simulations performed on the two SLs (figure SII).

| BTBZ03 Sample | Thickness (nm) | Roughness (nm) | Density (g/cm$^3$) |
|---|---|---|---|
| $BaTiO_3$ | 2.38 | 0.4 | 6.1 |
| $BaZrO_3$ | 4.19 | 0.4 | 6.5 |
| **BTBZ09 Sample** | | | |
| $BaTiO_3$ | 4.35 | 0.5 | 5.6 |
| $BaZrO_3$ | 6.75 | 0.5 | 6.23 |

Table 2: Lattices parameters, periodicity and thickness of layers of BTBZ03 and BTBZ09 SLs.

| Sample | $a_{SL}$(Å) | $d_{BZ}$(Å) | $d_{BT}$(Å) | $\Lambda$ (Å) | $e_{BZ}$(Å) | $e_{BT}$(Å) |
|---|---|---|---|---|---|---|
| BTBZO3 | 4.14±0.01 | 4.24 | 3.90 | 65.75 | 42.35 | 23.40 |
| BTBZO9 | 4.14 ± 0.01 | 4.22 | 3.86 | 110.97 | 80.09 | 30.88 |